\documentclass[a4paper]{article}
\usepackage[margin=2cm]{geometry}
\usepackage{graphicx}
\usepackage{lscape}
\usepackage{dcolumn}
\usepackage{bm}
\usepackage{amssymb}
\usepackage{stmaryrd}
\usepackage{fdsymbol}
\usepackage{xcolor}
\usepackage{cite}
\usepackage{authblk}
\usepackage{float}

\title{Thermal dependence of the mechanical properties of NiTiSn using first-principles calculations and high-pressure X-ray diffraction}
\author{P. Hermet}
\author{J. Haines}
\author{D. Granier}
\author{M. Tillard}
\author{P. Jund*}
\date{}
\affil{ICGM, Univ. Montpellier, CNRS, Montpellier, France. \par*\texttt{Philippe.Jund@umontpellier.fr}}

\begin{document}
\maketitle
\begin{abstract}
In this work we aim to study the effect of temperature on the mechanical properties of a solid. For this, we have introduced a new first-principles based methodology to obtain the thermal variation of the elastic constants of NiTiSn, a multifunctional Heusler compound. In parallel using X-ray diffraction, we have measured the isothermal bulk modulus at 300~K. The agreement between the calculations and the experiments is within the experimental error bars showing the accuracy of the calculations. Using two other numerical methods, which give all coherent results, we have shown that NiTiSn conserves its very good mechanical properties up to 1500~K. In particular at 700~K (the best working temperature for thermoelectric applications), NiTiSn remains a ductile and robust material making it a compound of choice for applications in which large temperature fluctuations are present.  
\end{abstract} 

\section{Introduction}
The impact of temperature on the mechanical properties of materials is often neglected. Indeed, most of the measurements are performed at room temperature and most of the calculations (in particular first-principles calculations) assume T=0K. Nevertheless in a number of real life or industrial applications, the temperature varies tremendously and thus knowing the resistance of a material becomes essential. Heusler and half-Heusler compounds are the topic of intensive research because they can be used in a great number of applications such as thermoelectrics \cite{shen}, optoelectronics \cite{kieven} or spintronics \cite{casper}. These compounds represent a class of ternary intermetallics associating three elements in the following
stoichiometric proportions 1:1:1 (half-Heusler) or 2:1:1 (full-Heusler). In this study, we will focus on the half-Heusler compound based on Nickel, Titanium and Tin atoms (NiTiSn), which is the archetype of half-Heusler materials for thermoelectric applications. It is a cubic crystal that crystallises in the F$\bar 4$3m space group with MgAgAs as prototype~\cite{pearson}. We already tackled in previous {\it ab initio} studies the electronic properties \cite{colinet}, the thermodynamic/mechanical \cite{pat} properties and the lattice thermal conductivity \cite{jalcom} of these compounds with a good agreement with available experimental results. Our main interest lies currently in the thermoelectric conversion of heat into electrical current via the Seebeck effect and a great deal of the scientific activity in this field is devoted to find materials with better conversion rates. Nevertheless, an important aspect in this quest is often ignored: even if a material is efficient, if it does not resist the large temperature variations involved in thermoelectricity it can finally not be used. Therefore we propose here a combined theoretical and experimental investigation of the thermal dependence of the mechanical properties of NiTiSn whose optimal working temperature for the Seebeck conversion is close to 700~K \cite{ZT}. A comprehensive study of the mechanical properties of half Heusler alloys has recently been undertaken by Rogl at al. \cite{Rogl} but it mostly involved experiments and more importantly the measurements were performed at room temperature with no investigation of the effect of temperature. The present work is mostly numerical but the measurement of the isothermal bulk modulus at room temperature (for the first time to the best of our knowledge) permits to validate the different methods to obtain its temperature dependence. In particular we propose an original numerical procedure linking the evolution of the elastic constants to that of the thermal expansion. This procedure can be useful to the community for the calculation of the mechanical properties of other materials.\\
For the sake of clarity we recall here the formalism permitting to obtain the elastic moduli once the elastic constants $C_{ij}$ have been obtained using the well known formulae for polycrystalline cubic materials with isotropic orientations (\cite{tasi} and references therein). The isothermal bulk modulus $B_T$ has been derived according to:
\begin{equation}
 B_T=\frac{1}{3}\left(C_{11}+2C_{12} \right).
\end{equation}
The shear modulus $G_T$ has been obtained taking the mean of the Reuss ($G_R$) and Voigt($G_V$) limits given by:
\begin{equation}
 G_R=\frac{5(C_{11} - C_{12})C_{44}}{4C_{44}+3\left(C_{11}-C_{12}\right)}.
\end{equation}
\begin{equation}
 G_V=\frac{C_{11} - C_{12} + 3C_{44}}{5}.
\end{equation}
Once these two moduli are obtained, the isothermal Young's modulus $E_T$ and the Poisson's ratio $\nu_T$ are obtained as:
\begin{equation}
 E_T=\frac{9B_TG_T}{3B_T+G_T}
\end{equation}

\begin{equation}
 \nu_T=\frac{3B_T-2G_T}{2(3B_T+G_T)}
\end{equation}

The paper is organized as follows: in section two and three the details of the calculations and of the experimental procedure are given. In section four we present our results and the major conclusions are drawn in section five.

\section{Computational details and method}

The isothermal elastic constants were calculated within the framework of the density functional theory as implemented in the ABINIT package~\cite{ABINIT}. The exchange-correlation energy functional was evaluated using the generalized gradient approximation (GGA) parametrized by Perdew, Burke and Ernzerhof~\cite{PBE}. The all-electron potentials were replaced by norm-conserving pseudopotentials generated according to the Troullier-Martins scheme. Ni($3d^8$, $4s^2$), Ti($3d^2$, $4s^2$) and Sn($5s^2$, $5p^2$)-electrons were considered as valence states. The electronic wave functions were expanded in plane-waves up to a kinetic energy cutoff of 65~Ha and integrals over the Brillouin zone were approximated by sums over a 8$\times$8$\times$8 mesh of special $k$-points according to the Monkhorst-Pack scheme~\cite{Monkhorst}. 

We used two formalisms to get the temperature dependence of the isothermal elastic constants $C_{ij}$. These two formalisms assume that the temperature dependence of the elastic constants are only caused by the thermal expansion ignoring anharmonic effects like the phonon-phonon interactions or fluctuations of the microscopic stress tensors.

The first formalism is the quasistatic approximation (QSA) and it requires a three step procedure. The first step consists in computing the elastic constants at 0 K as a function of volume using the density functional perturbation theory (DFPT) or the stress-strain method (here we preferred the DFPT). In the second step, the equilibrium volume V(T) at a given temperature T is calculated using the quasiharmonic approach. In the last step, the calculated elastic constants obtained from the first step at the volume V(T) are approximated as those at finite temperatures: $C_{ij}(T) = C_{ij}[V(T)]$. 

The second original formalism establishes a link between the temperature dependence of the elastic constants and the one of the thermal expansion. A convenient expression can be obtained as follows from the differentiation of $C_{ij}[V(T)]$:
\begin{equation}
dC_{ij} =\left( \frac{\partial C_{ij}}{\partial V}\right)_T\left( \frac{\partial V}{\partial T}\right)_P dT,
\end{equation}
where $P$ stands for pressure. Identifying the volume thermal expansion, 
\begin{equation}
 \alpha_V(T)= \frac{1}{V(T)}\left( \frac{\partial V}{\partial T}\right)_P,
\end{equation}
and taking the integration up to a temperature, $T'$, gives:
\begin{equation}
 C_{ij}(T')=C_{ij}(0)+ \int_0^{T'}\left( \frac{\partial C_{ij}}{\partial V}\right)_T V(T) \alpha_V(T)dT.
\end{equation}
According to this equation, the change of elastic constants with volume shows an almost linear behavior. We can therefore assume that a constant volume derivative is reasonable. In practice, we used the values of $\alpha_V(T)$ calculated combining the DFT and quasiharmonic approximation as reported in our previous work\cite{pat}. This method based on the integration of the thermal expansion will be called ITE in the following sections.
The isothermal bulk modulus $B_T$ was then derived from these two formalisms using the well known formulae for polycrystalline, isotropic cubic crystals (\cite{tasi} and references therein). Similarly the shear modulus $G_T$ was obtained taking the mean of the Reuss ($G_R$) and Voigt($G_V$) limits. Finally once these two moduli were obtained, the isothermal Young's modulus $E_T$ and the Poisson's ratio $\nu_T$ could be straightforwardly obtained.

\section{Experimental details}

NiTiSn was synthesized from a stoichiometric mixture of Ni (Sigma-Aldrich, 99.99 \%), Ti (Acros-Organics, 99.7 \%) and Sn (Alfa-Aesar, 99.85 \%). Powders of about 100 mesh were mechanically alloyed under pure Ar using Si$_3$N$_4$ grinding bowl and balls. Experiments were conducted at 400 rpm for 5 hours using a high-performance planetary Pulverisette 7 premium line Fritsch, GmbH and detailed in a previous work \cite{metals}. The powder was subsequently annealed under vacuum at 750$^\circ$C during 10 days. A conversion process from a multiphasic precursor into a single phase material is involved in this synthesis route leading to highly pure NiTiSn without residual tin. 

The NiTiSn powder was finely ground and passed through a 20~$\mu$m sieve. The powder was mixed with NaCl as a pressure calibrant using the equation of state of Birch \cite{Birch}. The mixture was loaded in a 200~$\mu$m diameter hole in a 75~$\mu$m thick tungsten gasket mounted on the anvil of a Merrill-Bassett diamond anvil cell (DAC). An ambient pressure measurement was performed prior to adding 4:1 methanol:ethanol as a pressure transmitting medium.

X-ray diffraction experiments were performed on a Bruker D8 Venture 4-circle diffractometer equipped with a Incoatec I$\mu$S 3.0 Mo microsource (110~$\mu$m beam, $\lambda_{K_\alpha}=$0.71073\AA) and a Photon II CPAD detector at a distance of 100~mm from the sample. The diffractometer was calibrated using a LaB$_6$ standard. The DAC was first centered optically, followed by centering along the beam direction by gasket shadowing. A rotation of 6$^\circ$ around $\omega$ was performed during acquisition. The acquisition time was 120~s. The diffraction images were integrated using the DIOPTAS program~\cite{Prescher}. LeBail refinements to obtain cell parameters were performed with FULLPROF~\cite{Rodriguez}. EOSfit7c was used to fit the data to a Birch-Murnaghan equation of state~\cite{Angel}.

\section{Results and discussion}
To obtain the variation of the mechanical properties of NiTiSn with temperature, the first step is to compute the variation of the three independent elastic constants for a cubic crystal: $C_{11}$, $C_{12}$ and $C_{44}$. They are shown in Figure 1 for temperatures up to 1500~K and obtained according to the two methods described in section III. The experimental $C_{11}$ and $C_{12}$ elastic constants measured by Rogl et al. \cite{Rogl} at room temperature are also reported in the figure and show a relative error below 8\% for $C_{11}$ with respect to our calculations. These experimental data on the NiTiSn elastic constants are the only ones reported in the literature. The first origin of this deviation could be the nature of the measurements. Indeed, we determined (in calculations and experiments) the isothermal elastic properties whereas Rogl et al. have measured the adiabatic ones. However, we calculated the adiabatic elastic constants from the isothermal one and we found that their difference is too small to be the main reason of this deviation. Furthermore, the absence of error bar in the reported experimental value by Rogl et al. makes the comparison with our results even more difficult. The second reason, the most probable for us, could come from the quality of the samples. As stated by the authors, the samples prepared using their multistep method, contained a certain number of secondary phases (Sn) and were polluted by Al or Si \cite{Rogl2} whereas our sample was remarkably pure as shown in Ref. \cite{metals}. 
\begin{figure}[H]
\begin{center}
\includegraphics[width=14cm]{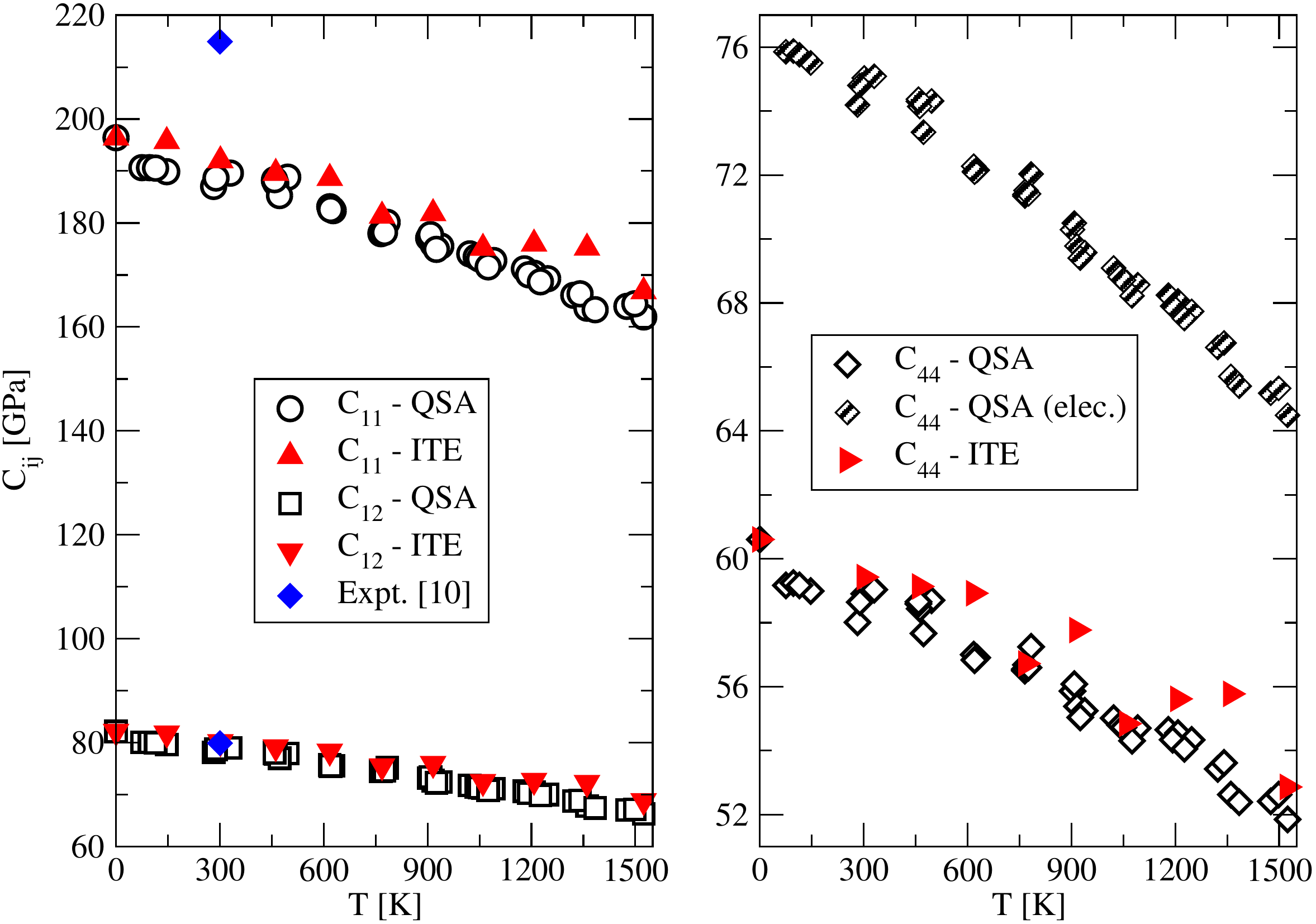}
\caption{Elastic constants vs temperature using QSA and ITE. In the right panel the contribution of the electrons to $C_{44}$ is compared to the total $C_{44}$. The reported experimental $C_{11}$ and $C_{12}$ elastic constants have been measured by Rogl et al. \cite{Rogl} at room temperature. We got the experimental $C_{12}$ from their article using $C_{12}=\frac{1}{2}(3B- C_{11})$.}
\end{center}
\end{figure}
The two sets of numerical methods reported in Figure 1 show the consistency of the theoretical descriptions and in particular the quality of the ITE method. We observe a relative small decrease of the elastic constants with increasing temperature over the whole temperature range (-15\% for $C_{11}$ at the most) for both theoretical models. This is a good indication that the material will keep its mechanical strength even at high temperature, a positive finding that is comforted by the small coefficient of thermal expansion found in our previous study \cite{pat}. An other interesting feature that can be seen in Figure 1 is the vibrational and electronic contribution to the elastic constants. Due to the local crystal symmetry, $C_{11}$ and $C_{12}$ only possess an electronic contribution which implies, according to Eq. 1, that the bulk modulus is only dependent on the electronic properties of the material. This is an interesting feature since by doping NiTiSn to improve the electronic properties for thermoelectric applications (Seebeck coefficient, electrical conductivity) one may also improve its bulk modulus. On the other hand the measurement of the bulk modulus in a collection of materials doped with different elements could give a hint of the material with the best electronic properties. The situation is different for $C_{44}$ since in this case it is possible to separate the electronic contribution from the vibrational one as shown in the right panel of Figure 1. It can be seen that if only the electronic contribution would be present, the solid would be much tougher with respect to a shear strain ($C_{44}$ is the main factor of the shear modulus). But the additional internal relaxation due to phonons allows some of the stress to be relieved, leading to a decrease of the overall $C_{44}$ by more than 20\%.

The elastic constants, $C_{ij}$, defined previously are obtained under the condition of a fixed (and vanishing) electric field $\mathcal{E}$: $C_{ij}=C_{ij}^\mathcal{E}$. The piezoelectric-stress tensor, $e_{\alpha i}$, can be indirectly determined from the elastic constants $C_{ij}^\mathcal{D}$, defined under the condition of a fixed (and vanishing) electric induction $\mathcal{D}$ according to:
\begin{equation}
C_{ij}^\mathcal{D}=C_{ij}^\mathcal{E}+\sum_{\alpha=1}^3\sum_{\beta=1}^3e_{\alpha i}[\varepsilon^{(\eta)}]^{-1}_{\alpha\beta}e_{\beta j},
\end{equation}
where $\varepsilon^{(\eta)}$ is the fixed-strain dielectric tensor. This situation occurs in thin films where a dielectric material is sandwiched between much thicker layers of other insulating host materials. In this case, the electrostatic boundary conditions fix the component of $\mathcal{D}$, and not $\mathcal{E}$, normal to the interfaces. In the case of NiTiSn, the independent elements of $C_{ij}^\mathcal{D}$ are:
\begin{eqnarray*}
C_{11}^\mathcal{D}=C_{11}^\mathcal{E} \\
C_{12}^\mathcal{D}=C_{12}^\mathcal{E} \\
\end{eqnarray*}
and
\begin{equation}
C_{44}^\mathcal{D}=C_{44}^\mathcal{E}+\frac{e_{14}^2}{\varepsilon^{(\eta)}_{11}}
\end{equation}
Only  $C_{44}^\mathcal{D}$ differs from  $C_{44}^\mathcal{E}$ and our calculated values for $C_{44}^\mathcal{D}$ are listed in Table 1 at 0, 300 and 750~K. $C_{44}^\mathcal{D}$ is always larger than $C_{44}^\mathcal{E}$ at all temperatures since the $e_{14}^2/\varepsilon^{(\eta)}_{11}$ ratio in Eq. 10 cannot be negative. The piezoelectric-stress constant has been derived and we obtain e$_{14}=$-0.714 C/m$^2$ at 0 K. This value increases by about 5\% at 750 K (see Table 1). At room temperature, e$_{14}=$-0.727 C/m$^2$. This value is about five times larger than the one reported in the most used piezoelectric material, $\alpha$-quartz \cite{Tarumi}. Nevertheless, although NiTiSn is stable within a large temperature range (no phase transition until decomposition), its low experimental electronic band gap ($E_g$ = 0.12 eV \cite{Aliev}) will strongly limit its use for piezoelectric applications (small short-circuit current).

\begin{table}[H]
\caption{Calculated elastic $C_{14}^\mathcal{D}$, piezoelectric-stress e$_{14}$ and fixed-strain dielectric $\varepsilon^{(\eta)}_{11}$ constants at 0, 300 and 750 K.}
\begin{center}
\begin{tabular} {lccc}
\hline
\hline 
      &    0 K   &300 K & 750 K  \\
\hline      
 C$_{44}^\mathcal{D}$ [GPa]         &  62.54   & 60.89 & 58.74  \\
 e$_{14}$ [C/m$^2$]                     &    -0.714 & -0.727 & -0.748  \\
 $\varepsilon^{(\eta)}_{11}$   &  29.70 & 30.09 & 30.67      \\
\hline
\hline
\end{tabular}
\end{center}
\end{table}

Once the standard elastic constants ($C_{ij}=C_{ij}^\mathcal{E}$) have been obtained, it becomes straightforward to determine the evolution with temperature of the mechanical properties starting with the variation of the isothermal bulk modulus shown in Figure 2 for the two theoretical approaches considered here.
\begin{figure}[H]
\begin{center}
\includegraphics[width=10cm]{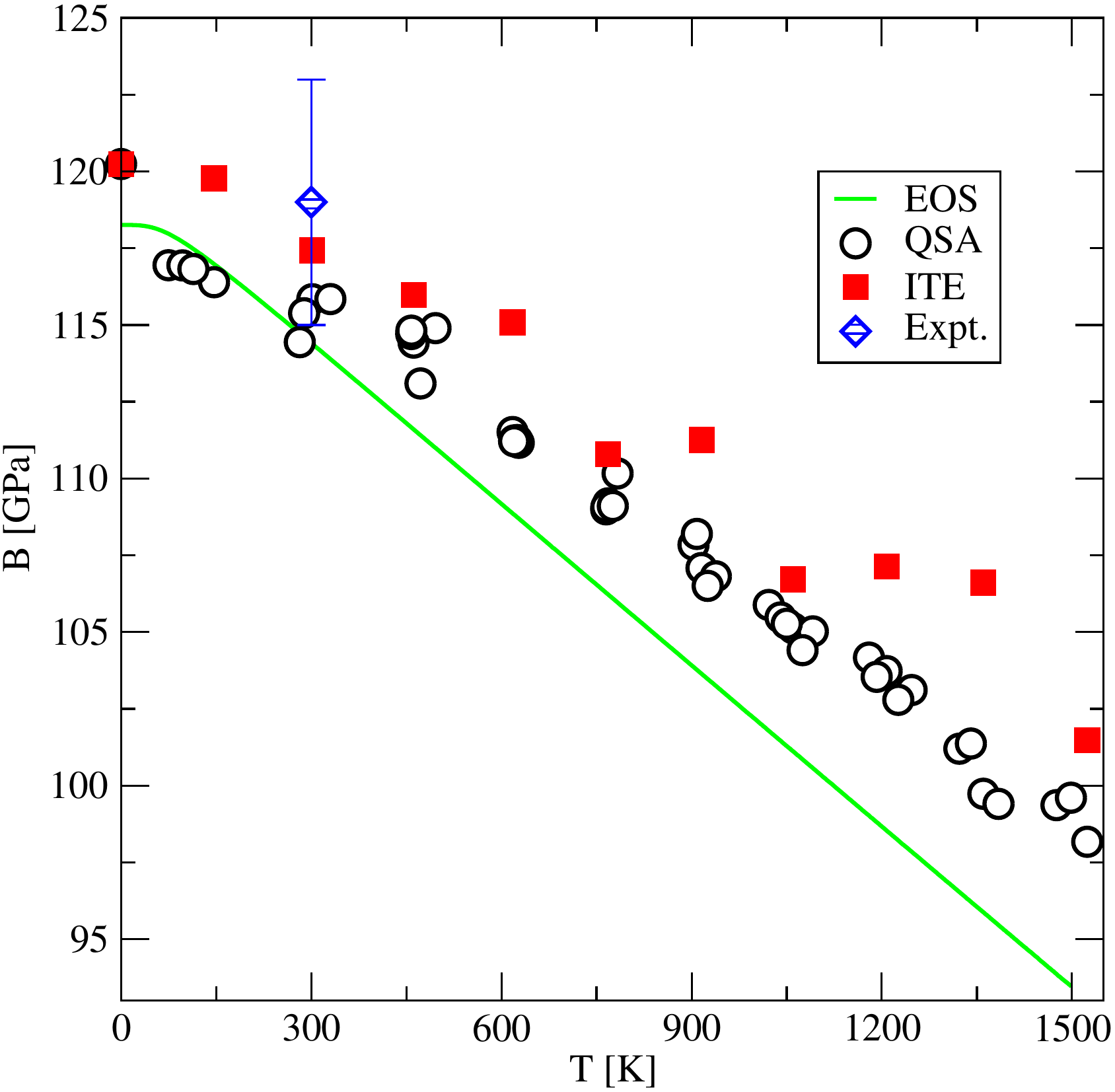}
\end{center}
\caption{Bulk modulus vs temperature: $\bigcirc$ QSA; {\color{red}$\blacksquare$} ITE; {\color{green}$\boldsymbol{-\!\!-}$} EOS;{\color{blue} $\meddiamond\!\!\!\!\!\!=$} experiment}
\end{figure}
In Figure 2 we have added the results obtained within the quasi-harmonic approximation after fitting the data to a Birch-Murnaghan equation of state (EOS) as obtained in our previous work \cite{pat}. The three models give consistent results with a relative error lower than 5\% at high temperature. Both ITE and the QSA give results at 300~K within the error bar (+/- 4GPa) of our experimental measurement showing the quality of these models.
This experimental value has been obtained from high-quality diffraction measurements from ambient pressure up to 7.5 GPa. Le Bail refinements (Fig. 3) were used to obtain the relative volume of NiTiSn (inset of Fig. 3) as a function of pressure. The relative volumes were fitted to a second-order Birch-Murnaghan equation of state yielding a bulk modulus of 119(4) GPa with an implied value of 4 for its first pressure derivative B'$_0$. This value is very close to the corresponding theoretical value, 4.5261. Expanding the equation to third order does not significantly improve the result.
\begin{figure}
\begin{center}
\includegraphics[width=12cm]{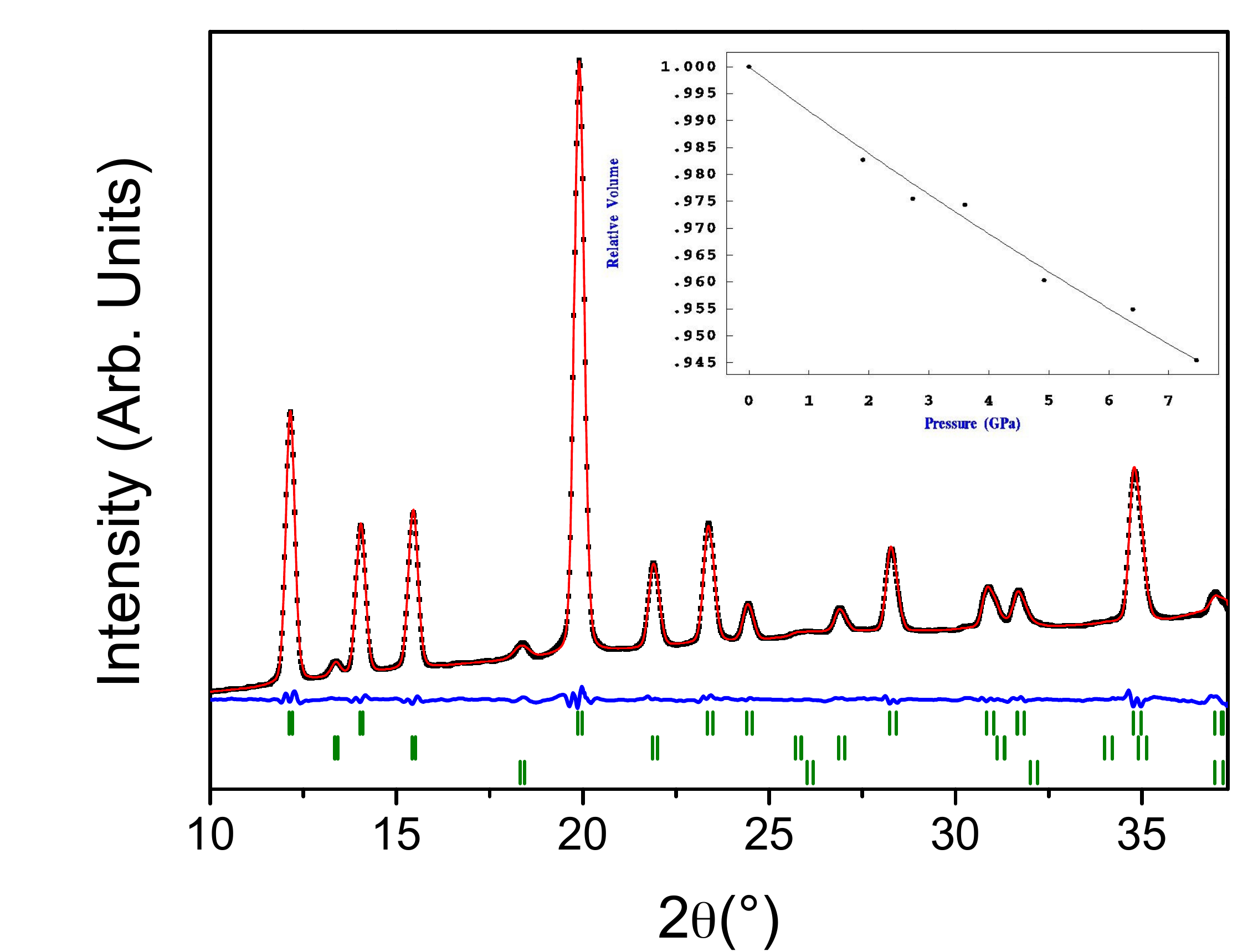}
\end{center}
\caption{Experimental (black), calculated (red) and difference (blue) profiles obtained from the LeBail refinement of NiTiSn at 7.5 GPa. Vertical bars indicate the calculated positions of the Bragg reflections (K$_{\alpha_1}$-K$_{\alpha_2}$ doublets, $\lambda_{K_\alpha}=$0.71073\AA) for NiTiSn (top), NaCl (middle) and W (bottom). Inset: Relative volume of NiTiSn as a function of pressure. The solid line represents a 2nd order Birch-Murnaghan equation of state with a B$_0$ of 119(4) GPa.}
\end{figure}
The fit obtained via the EOS gives globally smaller values than the two other approaches as well as the experimental determination but the general trend is similar. Our calculated and experimental bulk modulus at 300~K are lower than the values reported by Rogl et al. \cite{Rogl}. In particular, they reported a bulk modulus of 124.9 GPa whereas our measurements show a slightly smaller value of about 5\%. Nevertheless, this deviation remains acceptable considering the arguments previously given when comparing the elastic constants (Figure 1).

In order to close our investigation of the mechanical properties of NiTiSn as a function of temperature, we show in Figure 4 the evolution of the shear modulus, Young's modulus and Poisson's ratio which are generally the most used quantities to evaluate the strength of a material. Since the results obtained with QSA and ITE are similar, we have only reported the ITE results in Figure 4. 

\begin{figure}[H]
\begin{center}
\includegraphics[width=14cm]{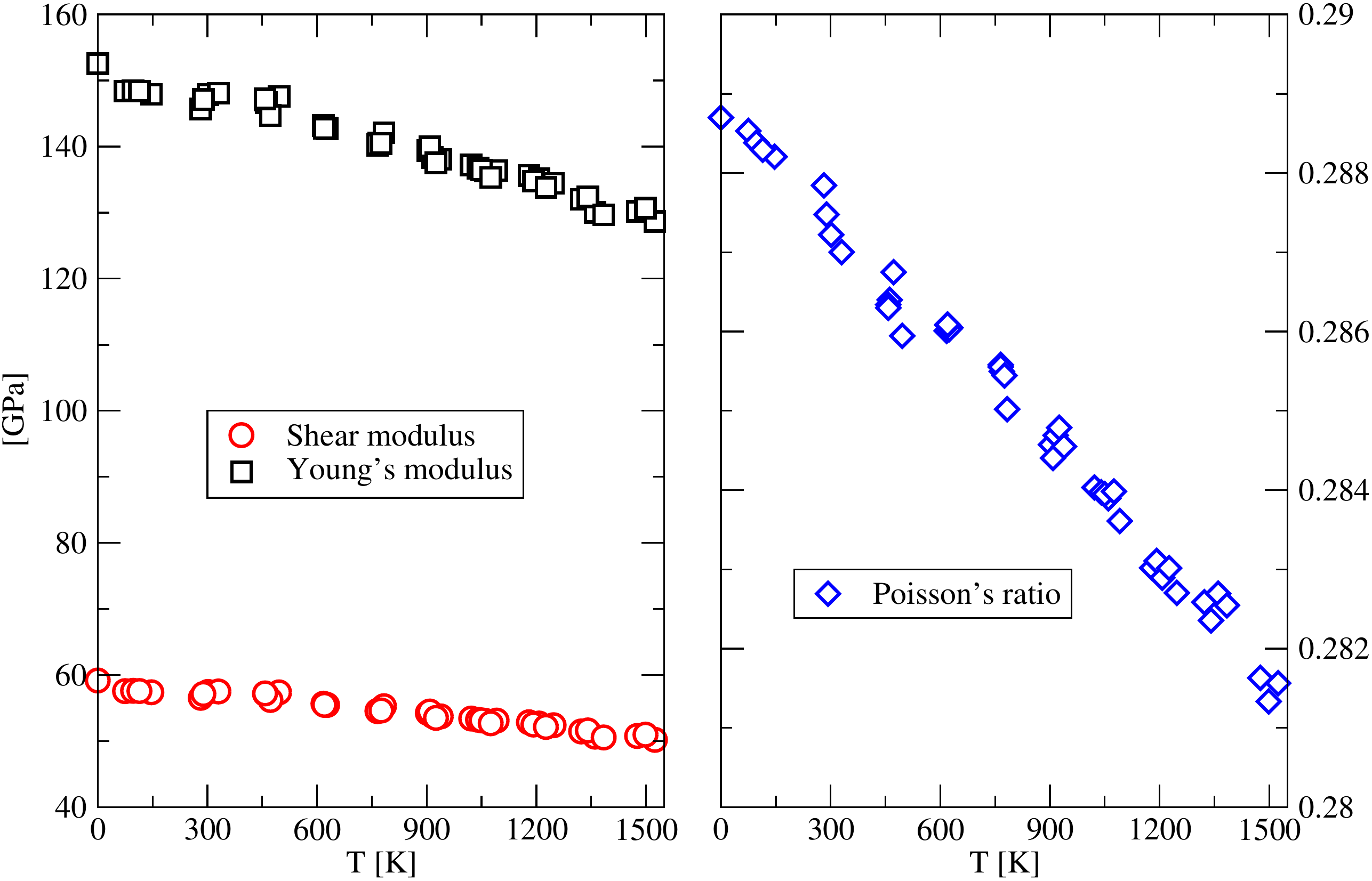}
\end{center}
\caption{Left panel: Shear and Young's modulus vs temperature; Right panel: Poisson's ratio vs temperature (results obtained with the ITE method)}
%
\end{figure}
As previously noticed for the bulk modulus, the other elastic moduli of NiTiSn decrease moderately as the temperature increases insuring a good mechanical stability of the material. According to the chart of the elastic properties of inorganic crystalline compounds (Figure 2 in Ref. \cite{chart}), NiTiSn possesses elastic properties above average and we show that this is still the case at high temperature. Pugh's ratio ($\frac{B_T}{G_T}$) is above 2 in the whole temperature range, showing a steady ductile behavior of NiTiSn. This is confirmed by the value of Poisson's ratio which decreases from 0.2887 at 0~K to 0.2815 at 1500~K. At 700~K which is the best working temperature for thermoelectric applications, Poisson's ratio is close to 0.285, a value comparable to the one of Tungsten \cite{W} or steel \cite{steel}. All these findings show that the half-Heusler NiTiSn has excellent mechanical properties whatever the temperature is. If we add the relatively small coefficient of thermal expansion \cite{pat}, our study shows that NiTiSn is an excellent material for many types of industrial applications involving large temperature gradients and in particular for thermoelectric applications.   

\section{Conclusion}

We have studied the evolution of the elastic properties of NiTiSn as a function of temperature with two numerical methods (three for the bulk modulus) including one based on the integration of the elastic constants that is completely original. In addition, high pressure XRD experiments were performed at 300K to obtain for the first time the isothermal bulk modulus. At that temperature, the agreement between the experimental measurement and the three numerical methods is within the experimental error bars showing the quality of the calculations. In addition using the elastic constants defined under a fixed electric induction, we show that NiTiSn exhibits at 300~K calculated piezoelectric properties five times better than $\alpha$-quartz.\\
Concerning the evolution of the mechanical properties of NiTiSn with temperature, we have shown that this material has above average properties in the whole temperature range (0-1500K) and remains ductile even at low temperatures making it thus easy to engineer. This remarkable stability of the elastic properties makes it an ideal candidate for thermoelectric applications where large temperature gradients are used. The numerical methods we have introduced in this work can be used by a large scientific community who wishes to predict the mechanical stability of a material with temperature, stability that has so far rarely been addressed apart within the quasi harmonic approximation.


\newpage



\end{document}